# Privacy & Social Media in the Context of the Arab Gulf

Norah Abokhodair∗
University of Washington
Seattle, USA
noraha@uw.edu

Sarah Vieweg
Qatar Computing Research Institute
Doha, Qatar
svieweg@qf.org.qa

## ABSTRACT
Theories of privacy and how it relates to the use of Information Communication Technology (ICT) have been a topic of research for decades. However, little attention has been paid to the perception of privacy from the perspective of technology users in the Middle East. In this paper, we delve into interpretations of privacy from the approach of Arab Gulf citizens. We consider how privacy is practiced and understood in technology-mediated environments among this population, paying particular attention to the role of Islam and cultural traditions in constructing norms around privacy. We then offer culturally sensitive design principles and suggestions for future research that incorporates previously unexplored characteristics of privacy, which play a role in how users navigate social media.

## ACM Classification Keywords
H.5.m. Information interfaces and presentation (e.g., HCI): Miscellaneous.

## INTRODUCTION
Layla is a 23-year-old Saudi girl. After finishing her undergraduate degree at King Saud University in Riyadh, she started working at a multinational consulting firm in her hometown. This particular firm is one of the few workplaces in the country in which gender segregation laws are relaxed and Layla can attend meetings with men. After a few months, Layla started receiving friend requests on Facebook from male colleagues. While she does not mind being friends with males on Facebook, she is aware that it would disturb her family. In the given situation, Layla was faced with a conundrum: should she placate her family and follow their wishes, or participate in social networking that may prove useful in her professional life?

Previously, Layla used her Facebook account to keep in touch with female friends and family. Maintaining a small group of friends on Facebook was culturally acceptable and safe. However, if she is to immerse herself in the culture of her new workplace, make friends, and advance her career, she may have to take on the task of managing her social media sites in ways that satisfy both the traditional expectations placed on her by her family, and the professional expectations placed on her by her colleagues.

## Motivation
Social Media Sites are playing an increasingly important—yet frequently problematic—role in the everyday lives of users in the GCC[1]. They bring about challenges in terms of the ways this population experiences privacy and manages identity. We are aware of various conceptions of privacy that exist across cultures, but when it comes to commonly used tools such as social media sites and microblogging services, the assumptions around privacy that are embedded in technology design do not necessarily make it easy for those who adhere to Islamic interpretations of privacy to use them. In other words, we can say that a particular interpretation of privacy is foisted upon the users of these technologies, and if those users regard and practice privacy differently, then re-use and appropriation must take place. In this paper, we focus on how social media users in the Arab Gulf manage their interactions, and their identities, while maintaining traditional expectations and practicing privacy in culturally acceptable ways.

### Privacy, Religion and HCI
Researchers have long discussed, researched, and debated definitions of privacy and how it should be considered and incorporated into technology design. Recently, scholars have questioned the notion of a universal definition of privacy, suggesting instead that it is a socially constructed and culturally bound concept [10,29,31]. While various outcomes and applications have arisen from the studies to date, limited research approaches privacy from a point of view that does not focus on interpersonal boundaries [9], but allows for alternative interpretations of "privacy."

For technology users from the Arab Gulf, the Arabic language, religious affiliation, and cultural expectations are factors that play a critical role in technology adoption and use. In particular, understandings of privacy are tied to expectations and norms that have foundations in Muslim religious practice. The importance of privacy is borne of the responsibility to maintain the sanctity of one's body and one's home, in addition to upholding the honor and good name of one's extended family.

---

∗ This research was conducted while an intern at Qatar Computing Research Institute (QCRI)

[1]The GCC stands for the Gulf Cooperation Council—six Muslim majority countries near or on the Arabian Gulf that share cultural and historical ties.

Researchers have noted ways in which Muslim social media users enact norms that dictate modesty and consider the importance of reputation, self-image and family honor, [1,7,34,44], but a focus on privacy vis-à-vis social media use has yet to be theorized. Looking at privacy through the lens of digital technology in an Arab Gulf setting raises provocative questions about the ways in which cultural norms and expectations that are rooted in tradition come to pass when they are interpreted via modern media.

Our aim is to contribute to the literature on privacy in two ways. First, introduce and explain privacy vis-à-vis social media use, with an eye toward Arab Gulf understandings and enactments of privacy. Second, we provide evidence regarding how privacy is enacted in digital environments through analysis of face-to-face interviews. Combined, these serve as a foundation for design opportunities that take a more encompassing view of privacy.

**RELATED WORK**

Previous research contributes to understandings of privacy as it is understood in the Arab Gulf, but questions—and the need for additional theorization regarding privacy—remain. Privacy is often viewed as a concept of the self as an individual, apart from a group [9]. While this is true in many previously studied user groups, we note that in the Arab Gulf the basic idea of *the self* and *the group* is conceptualized such that asserting one's individuality is viewed in a negative light. Membership in a family and tribe are of the utmost importance; there is no individual separate from a family [12,20]. This perception of the self extends elsewhere, as explained by Bidwell regarding her research on Facebook use in rural Africa: "In the everyday acts, linguistics, and philosophy of precolonial African society, personhood was constituted by others—both living and dead—and identity was perceived as undifferentiated and fixed to a social position" [12]. This idea of formulating the self through others is a process that applies also to Arab Gulf culture, and as this relates to privacy, the notion of "personal privacy" takes on a different tenor.

**Privacy in Design**

Scholars have also noted how conceptions of privacy become inscribed in technology design [7, 17, 22, 23, 43]. As the design and HCI communities strive to establish a broader understanding of values imbued in technology design, there is risk in continuing to omit alternative perspectives toward privacy. We strongly support the idea that "A first step toward designing for privacy entails understanding what privacy means to those who will use and be affected by the use of the technology" [23].

This is not to say that researchers and scholars have not carefully considered the highly contextual, culturally-bound nature of privacy within the realm of technology use. Palen and Dourish [31] delve into the ways in which technology is viewed and implemented vis-à-vis cultural norms. The authors consider privacy through the lens of Altman, which leaves room for us to interpret what is meant by "privacy," and not be beholden to a particular perspective. Altman present privacy as a dynamic, dialectic process "conditioned by our own expectations and experiences, and by those of others with whom we interact" [9 in 31]. Building on their work, Barkhuus emphasizes that "privacy is not an easily measurable unit…we as HCI researchers and practitioners need to approach the notion through more contextually grounded measures" [11].

However, while many have thoughtfully analyzed and theorized privacy and technology use, there has been little research that brings Gulf Arab notions of privacy into the conversation. One study that does focus specifically on privacy and social networking in the GCC is by Faisal et al. [20]. They evaluate privacy behaviors, trust concerns, and attitudes of social media users in Kuwait. The study included 222 surveys, with results showing that a high level of privacy awareness and trust are major factors that affect how Kuwaiti youth participate on social media. In addition, they note that Kuwaitis "tend to be conservative and favor collective over individual interests." Overall, they stress that privacy is a high-stakes value in collectivist and honor-based societies, and a violation of privacy "leads to shame and loss of face" [20].

Contributing to the body of research on privacy enactment via digital technology among GCC users, and answering the call made by [11] for "a more nuanced treatment of the notion of privacy" we add to the conversation around Gulf Arab interpretations of privacy. We provide a description of privacy from the perspective of citizens of the region, and go on to elucidate how the enactment of privacy affects their behavior in digital environments.

**UNDERSTANDING PRIVACY IN THE GCC**

In GCC countries, knowledge and practices around morality and ethics are primarily derived from Qur'anic text, and the example (*Sunnah*) of the prophet Muhammad. Together, these constitute the principle sources of *sharia law*, which is the foundation for both the judicial system, and societal norms and expectations regarding behavior. The need for, and expectation of, privacy plays a prominent role in daily life. To adequately grasp how privacy is viewed in the GCC, it is important to first explain what privacy is from an Islamic perspective. We do so by providing background on how privacy is viewed in the Arab Gulf. We then focus on the Islamic roots, and discuss why the consistent enactment of privacy is tantamount to following sharia law.

While prevalent conceptions of privacy focus on the individual, and their relationship to "a group," (namely, notions of seclusion and control) [9] Gulf Arab notions of privacy are bound up in the importance of modesty. [37] explains that "the notion of privacy in the Arab-Islamic paradigm is largely related to the requirement of modest self-presentation for Muslims in public, particularly women…the underlying meaning of privacy in the Arab-Islamic culture is respect and not seclusion." Presenting oneself as modest, in both dress and behavior, is of great importance to being a respectable member of Gulf society, and privacy plays an important role in how modesty and respect are maintained.

Research that focuses on Arab notions of privacy comes from fields including healthcare, marketing, and architecture. Drawing on experiences with the psychiatric care of Arab patients in the United States, [27] note that privacy is vigorously guarded, and medical personnel are often viewed as intrusive when asking questions that Western patients consider routine. It is not uncommon for Arab patients to answer questions in a way that pleases the medical professional, but not necessarily be honest about a condition or symptom, so as to maintain privacy [27].

From an architectural perspective, Sobh and Belk provide detail about traditional home design in the Gulf Arab countries, where houses are typically designed with an inward-facing center to protect the family from the public eye [37]. The idea is to maintain the sanctity of the home, which is considered sacred and pure, and which must be guarded from the gaze or intrusion of non-family members.

When it comes to marketing, particularly of Western products/ideals in the Arab Gulf, Sobh and Belk again point to slippage in current approaches that fail to understand the importance of privacy. For example, some advertisements show gatherings of mixed-gender, non-family members in a home, which is forbidden in Gulf Arab culture. The privacy of females in particular is of the upmost importance, and the idea of a mixed-gender gathering is nonsensical [37].

**The Islamic vision of privacy**
To elaborate on these approaches toward privacy, we must delve into the significance privacy has in the GCC. The way in which privacy translates so as to be understood by a non-Gulf Arab audience involves several concepts.

"Hurma" (حرمة) is an Arabic word that symbolizes the concept closest to the notion of privacy in the English vocabulary. Depending on the context, hurma has two meanings: 1) anything that is unlawful to obtain or look at without permission or 2) a woman, a sacred space (mosque or home) or a sacred time (holy month). These are all considered pure, and should remain guarded; any intrusion on their sanctity is considered sinful [18].

In the Quran, privacy is first mentioned in the context of instructing people to seek permission before entering another's home. The purpose is to protect the sanctity—or hurma—of the house and the body. One is required to knock on a door three times before accessing another's space. This rule is in place to avoid walking in on another while in a state of undress, or while with one's spouse/family. Entering without permission risks exposing one's "awrah"(عورة). In Islam, awrah literally means the intimate parts of one's body. In the GCC, everyone is instructed to cover certain body parts; what is covered and when depends on the specific situation and the level of conservativeness. For women in the GCC, awrah can extend beyond the hair to include the arms, the legs, and in some cases the face. Awrah and hurma can be used together to indicate the unlawfulness of obtaining or seeking access to others' sacred body parts and spaces.

In addition, Saudi Arabia upholds the right to privacy in its Basic Law of Government: "Dwellings are inviolate. Access is prohibited without their owners' permission. No search may be made except in cases specified by the Law." [39]. In the GCC, the rule associated with privacy is "haq al-khososyah"(حق الخصوصية), which is an "individual's right to protect some aspects of their private life and maintain confidentiality to safeguard his/her reputation and aspects of his/her life that are kept away from the interference of people" [5]. When GCC citizens protect their hurma and/or awrah, they are enacting their right to haq al-khososyah.

We suggest that privacy as practiced in traditional Islamic cultures, especially in the Gulf, involves three aspects; awrah, hurma, and haq al-khososyah. We can think of awrah as an object, something tangible that must be protected. Hurma is the value, borne of Islamic teachings that asserts the sacredness of awrah. Haq al-khososyah is the law that guards and protects one's right to invoke hurma in the interest of protecting awrah. In Figure 1, we depict awrah at the center, as it is of the most value; it is the object to be shielded. We show hurma encompassing awrah; it is the space that surrounds the awrah, protecting it. Haq al-khososyah encircles both hurma and awrah; it empowers people to legitimately protect their awrah.

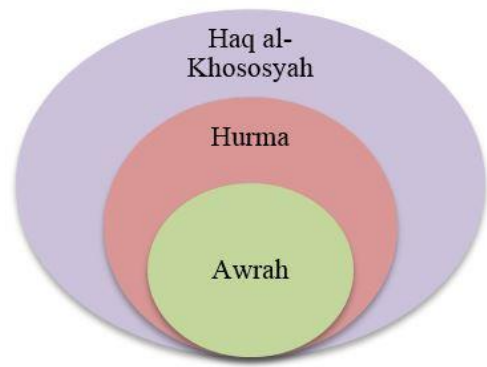

**Figure 1: Islamic Interpretation of 3 aspects of "privacy"**

Privacy is again mentioned in the Quran in relation to a law that promotes the respect of others by warning Muslims to refrain from bad manners that can lead to an invasion of others' privacy. This is based on the prohibition of spying, gossiping, and mistrust. When a rule is introduced in a holy text, Muslims are obligated to follow it. Failure to follow these rules is considered a sin. Therefore, respecting others' privacy, and maintaining one's own privacy, is a decree from God. GCC citizens' practice of and adherence to privacy as part of their daily lives stems from the pursuit of the blessings of God, or "barakah" (بركة). This endeavor leads to the creation and maintenance of a respectful image.

**THE STUDY**
The population we focus on are citizens of two of the six Gulf Cooperation Council (GCC) countries—Saudi Arabia, and Qatar. The GCC is a political and economic union that implements similar regulations in trade, finance, and religion (among other matters), and whose governments and leadership maintain close ties with each other. Islamic principles and tribal traditions are thoroughly intertwined and provide the foundation for GCC customs, laws and practices. Virtually all citizens of the GCC are practicing Muslims; the majority adheres to conservative interpretations of Islamic texts which dictate their daily lives. Arabic is the official language of the GCC, and the language of the Quran; it is the primary language taught at public schools and used in official government processes. However, English is commonly spoken; it is the second most common language among citizens of the GCC and is used in many day-to-day interactions. In addition, English classes are compulsory in schools.

GCC society is collectivistic in nature. Patriarchal norms direct behavior; the reputation of one's family and tribe are of the utmost importance and take precedence over individual status, needs and/or self-image. In collectivistic cultures, individuals identify as part of strong and cohesive long-term commitments to a whole [42] that can be seen in the relationship people have with their immediate and extended family, and their tribe. In the Quran, the practice of maintaining a relationship with kin is referred to as 'silatur-rahim' (maintaining the bonds of kinship) meaning one must "deal properly with relatives, supporting them with whatever possible and warding off bad things. "[It involves] visiting the relatives, asking about them, checking on them, giving them gifts when possible, helping their poor members, visiting their sick members, accepting their invitations, having them as guests, feeling proud of them and elevating them" [35 in 2]. In Qatar and Saudi Arabia, it is a regular practice to allocate some time during the weekend and the holidays for visits between relatives. This practice is highly encouraged in Islam and there are verses in the Quran that warns people from severing or abandoning these ties.

Additionally, citizens of these countries follow strict social conventions akin to unspoken rules. For example, gender segregation is an important aspect of life in the GCC. This practice concerns the interaction between women and men who are not close blood relatives (non-mahram[2]). Gender segregation is actively enforced in Saudi Arabia and commonly practiced in Qatar (though it is not legally required). This segregation is in place "to regulate women [and to] prevent other men from encroaching on the male honor of the family" [3]. When in public, women must wear loose robes over their clothing called *abaya*, cover their hair and sometimes their faces with a *shayla* and *niqab* respectively, all to avoid the "unwanted gaze" of unrelated men. These are everyday practices borne of an adherence to modesty and respect. However, women are not expected to wear the shayla and abaya when in the company of other women, nor around mahrams.

Citizens of the GCC are also avid users of social media. Recently, Saudi Arabia was ranked second globally and first among Arab countries for using Snapchat [8]. Additionally, in 2014, the kingdom also recorded a rate of over 7.8M active Facebook users, as well as the fastest-growing population on Twitter with more than 5M active users and 150M tweets per month [38]. Numbers are similarly high in Qatar: 65% of Qataris are on Instagram, 44% are on Facebook, and 46% are on Twitter [32]. Fast adoption, high rates and regular use of social media will continue to grow in the region. However, despite the popularity of social media, users find it necessary to employ workarounds that allow them to use these technologies in culturally appropriate ways.

Privacy is a "multifaceted" notion [15]; it is interwoven with notions of trust, identity, seclusion and autonomy. In the Arab World, and in the GCC in particular, honor, shame, and modesty are also factors that influence how privacy is approached and carried out. We are interested in how "privacy" as practiced in the Gulf is enacted through the use of technologies that were not developed with this user group in mind. We asked participants about their uses and habits regarding social media, and talked about their opinions and preferences. In analyzing interview data, it became clear that privacy was a concern.

**Method**
The authors have both been actively conducting research in The GCC The first author is a Saudi Arabian female, the second author is an American female who has lived and worked in Qatar for several years. Each of us has previously interviewed citizens of the region about their views and habits regarding social media use. Through this ongoing research, we came to realize that when discussing social media use, privacy was a topic that was frequently discussed among all participants.

In embarking upon joint research that considers privacy in a Gulf Arab context, we found ourselves constrained by definitions and conceptions of privacy that are based in a perspective that is largely theorized and defined in American and Western European perspectives. These perspectives do not translate in a way that allows for "privacy" in a Gulf Arab sense to be fully acknowledged or understood. This challenge is not unique to this research, and we have observed that other researchers who approached the topic of privacy in the Arab Gulf often juxtapose their explanations of privacy with "Western" definitions [1, 2, 20, 37].

---
[2] A *mahram* is a male relative that a Muslim woman cannot marry, or to whom she is already married, e.g. father, husband, brother, son.

*Participants*

The first author focuses on Saudi Arabia, and the second on Qatar. Each recruited participants through personal and professional connections, who then recommended additional participants. We used a "snowball" sampling approach due to the difficulty of recruiting research participants in the GCC. It is challenging to find people to participate in qualitative research due to the perception that discussing one's personal life with a relative stranger is an invasion of privacy. For many Arab people, discussing personal matters is uncomfortable and typically avoided, especially outside of the family. In fact, Al-Munajjed [3] states, "a researcher is not readily accepted in a traditional milieu that frowns upon those enquiring into other peoples' lives." In addition, in Saudi Arabia, gender segregation is widely practiced, and recruiting willing participants requires using personal connections to find people who will sit down with someone not of their own gender.

*Interviews*

Our sample for this research is comprised of thirty-three face-to-face interviews with eleven men and twenty-two women that lasted from 21 minutes to 2 hours. We conducted interviews in English and Arabic between August 2014 and November 2015. Participants are citizens of either Saudi Arabia or Qatar, at least 18 years old, and all have at least some college education. Participants are native speakers of Arabic, and are fluent or bilingual in English.

The first author is bilingual in Arabic and English, and asked participants for their preferred language. At the outset, most preferred English, though code-switching and borrowing occurred. When discussing sensitive topics, the first author found that participants tended toward English when they spoke about experiences grounded in the time they spent in the US and Western Europe. In addition, based on our observations and personal experience, there is a sense that speaking English is "cool," and young people like to speak English when given the opportunity. The second author conducted all interviews in English. Before we met participants, we told them we would speak about general social media use. We started the interviews with semi-structured questions such as "What social networking sites do you use? What do you like/dislike about them? Do you wish anything were different about a particular site? What challenges do you face in using social media?" These general questions led to free-flowing conversations that were often confessional in nature. Participants spoke about social media use, but also shared secrets about culturally unaccepted activities such as dating and attending mixed-gender gatherings.

Overall, we do not claim that our sample is representative of either country, nor the region. In general, in conducting this interpretive/descriptive research, our aim is to settle upon "a specification of the characteristics of the sample, so that one can make judgment about the applicability of the findings" [19]. I.e. our goal was to tap into this population, identify themes in the data, and use those themes as the foundation for culturally aware design principles.

**Data Analysis**

Authors transcribed their own interviews, paying close attention to paralinguistic cues such as tone, pauses and laughter. In addition, when Arabic was spoken, the first author took context into consideration, i.e. she was careful to preserve the meaning, not simply the syntax of the interview. We then performed open coding on the transcripts to generate high-level themes. This step involved an iterative process that generated specific themes from more general themes [40]. Additionally, we conducted action-implicative discourse analysis [40]—focusing on mentions of privacy—to gain insight into how privacy was interpreted, enacted, and framed vis-à-vis social media use.

*Analyzing Interview Data*

We draw from a range of analytic methods to understand and interpret our data. First, we draw from action-implicative discourse analysis (AIDA) [40,41]. Performing AIDA first involves transcribing the text, and capturing words, vocal sounds (e.g., um, hmm), and repetitions. Next, analysts use theoretically informed induction, in which "tapes and transcripts are repeatedly studied to identify interesting practices, where notions of what is interesting are shaped by knowing what observed practices would challenge or extend theorizing" [40]. In our study, we focus not only on the interview transcripts, but also on observations, and background knowledge of the context and culture. This type of analysis focuses on language use as part of the participant's lived experience, i.e. we look at what the interviewees are saying within the context of the Arab Gulf culture. This allows us to understand participants' discourse as an information source regarding their beliefs and views.

Being "cultural insiders" is an advantage in this situation as it allows us to better explicate and explain the values as commonly practiced in Gulf societies. We generated themes based on our readings of the whole group and triangulated results with each other throughout our analysis.

**FINDINGS**

**On the meaning of "Privacy"**

Readers will note that throughout this paper, we have not provided a definition for "privacy." This is due to our commitment to understand privacy based on the views and enactments of our population [22], and to consider it as an ever-evolving, locally-constructed phenomenon. Our approach to framing privacy stems from referring to Nissenbaum's framework of 'contextual integrity,' which is concerned with the norms that regulate the flow of certain types of information [29]. We rely upon contextual integrity as the benchmark for how we understand privacy in the context of our user group. Based on an analysis of our interviews, we see how privacy is about more than managing personal boundaries. When asked about their views of privacy, participants often referred to societal expectations rather than personal limits or beliefs. They spoke of an ongoing boundary negotiation that encompassed their entire families, and how they act as representatives to meet societal expectations. This is illustrated in a male Saudi's (M3) response to a question regarding privacy:

"It [privacy] is not about me and my beliefs; it is about the audience and what they believe in and their objective."

M3 is very clear in his assertion that in the case of a Saudi audience, the maintenance of privacy is not related to what he believes is permissible or ethical, but is instead driven by societal expectations.

A Saudi view of privacy is expanded upon by (F5), a female who spent a year living in London studying for her master's degree, and who has since returned to Saudi Arabia. When asked about her sense of privacy in Saudi compared to London, she said:

*"Privacy is a lot more of a right over here [in Saudi] than anywhere else. People actually respect your privacy here, where in the West you will be asked too many details. Generally speaking, the idea of privacy in the West is hard to achieve...sometimes it is something you have to stand up for. Like if you want to buy something at the pharmacy there [London] they ask you for a photo ID. They don't ask for one over here unless it is something super serious."*

To F5, the benign request for ID at a pharmacy—which is not a common practice in Saudi—to purchase over-the-counter medication, is seen as a violation of her privacy. She explains that in Saudi Arabia privacy is not something that must be actively pursued, because it is a given, communal, expected value that is enacted through lifestyle, local norms, expectations, and laws.

This idea of privacy as a communal effort is elucidated by (F7), a Saudi female:

*"Privacy is important but it is not in my hand…I wouldn't say it is a need."*

The phrase "it is not in my hand" refers to the understanding among citizens of the GCC that privacy is achieved by a group; individuals do not approach privacy with the goal of protecting only themselves. In effect, F7 is saying that privacy is a collective practice that she alone would not be able to achieve.

*Private accounts*
In our sample most participants (especially females) maintain private social media accounts. In the below excerpt, a Saudi male, (M8), explains the tension between wanting to be online—which entails conforming with the rules of the platform—and the need to adhere to social expectations, which dictate that private accounts are the default choice. The key challenge for GCC users is negotiating modernity while maintaining their authenticity and tradition. In other words, the challenge is *managing transcendence* and coherence in the age of social media [1].

In explaining the reason she had to make her Facebook account private, (F4), a female Saudi, says: "[B]ecause of the community and because of the environment. Personally I don't find something wrong about being a public person [on social media] but if you don't comply necessarily with all the rules around you in your environment, then sometimes it is not a good idea."

Overall, the option to have a private social media account is not something that is always necessarily chosen because a user wants to maintain a private account, but because if they do *not* have a private account, the societal and familial backlash they would likely encounter is too great a risk. Privacy is not personal; it is the foundation for how entire families and tribes uphold their reputations.

*Information Disclosure and Boundary Regulation*
In his writing on privacy regulation theory, Altman argues against a definition of privacy that relies upon the idea of the withdrawal of an individual from the group. Instead, he advocates for an understanding of privacy as a process of optimization [9]. This interpretation provides a nuanced way in which to understand the privacy behavior of social media users in the Gulf. The ways users find the right balance between withdrawal and disclosure is very much tied to the idea of optimization; figuring out how, within the expectations of their culture, can they successfully use social media to share, communicate and have fun, while at the same time, maintain family honor?

Other social media researchers have studied the relationship between privacy concerns and the level of information disclosure and boundary regulation as a protective measure. Marwick and boyd [26] discuss tactics Twitter users adopt to navigate multiple audiences while avoiding "context collapse," or the divulging of information to the wrong audience. In addition, Farnham and Churchill [21] provide an in-depth discussion around issues of "the singularity of identity" on social media. The authors assert that there are many consequences for the user from the assumption that technology users have a unique, all-encompassing, solitary identity. They assert that "in reality, people's lives are 'faceted'; that is, people maintain social boundaries and show different facets" depending on the situation.

Our interviews support the claims of [26] and [21]. We heard about the acquisition and management of multiple accounts that reflect different personas on Facebook or other sites, the maintenance of accounts on different social media for different audiences and purposes (e.g. maintaining an account on Twitter for work-related posts, and Facebook for socializing with friends); the use of pseudonyms, and the management of boundaries between friends' circles and the information they could see online.

Participants also talk about different degrees of information disclosure on social media, which depend on both audience and platform. For example, a user may want non-Saudi friends to see a particular post or photo, but not Saudi friends nor family. Or, a female may want to show a photo of herself without her shayla to only female friends and mahrams. In addition, it is important to note that images carry far more weight, and risk, than text. If a photo or video is spread outside a user's intended audience, it can bring great shame and embarrassment to the user and their family. In some cases, the inappropriate sharing of an image can severely threaten reputation, which has far-reaching consequences on livelihood, including marriage prospects, career opportunity, and general well-being.

To illustrate the significance of photos, below is an excerpt from (F6), Saudi female, in which she shares her concern regarding her father gaining access to her account:

*"I have my dad blocked on Facebook (laughing). The moment I learned he opened an account I blocked him because at the time he opened it I was getting tagged on a lot of pictures that I didn't want my family to see. Pictures of traveling photos or wearing shorts, etc.) I didn't even want to risk (loud voice) something accidently being set on the wrong privacy setting. Even though I always had my privacy setting on pretty high. The only things I allow to be shown are the things I allow. I have ten different friends groups. I have 'limited', 'more limited' 'most limited'…I have six to seven different levels of privacy."*

F6 goes through great lengths to make sure that her Facebook account is finely tuned to ensure particular information is shared only with the chosen audience(s). If her father or other family members were to see that she has photos in circulation that show her wearing shorts for example, her reputation would be severely damaged, and her family's honor would be brought into question.

Regarding the preservation of family honor, a male Saudi, M3 describes his privacy settings on Facebook:

*"It is four layers. I am an old school Facebook (illustrates his layers on paper). To me, basically, you have a big circle here which have the pictures and the tags, the tags, the tags (he is repeating them to stress their importance)…the tags are full of details on college life and many things, so I don't want to have anyone seeing them."*

M3 stresses that while he is an avid user of Facebook, and had already configured his privacy settings to be quite strict, there are still things he cannot control, such as tagging by other users. Features on social media sites—such as the "like" and "tagging" features on Facebook—can have unintended consequences for users who are in a position to have their image put at risk by the accidental sharing of photos or information. Females in particular are in danger of exposure, as they bear great responsibility in upholding their family's reputation [36].

(M1), a male Saudi, describes some of these concerns:

*"Four years ago when Facebook changed the privacy settings, one of my friends liked a picture of my family, the picture had my sister in it, and I got really upset and embarrassed that he could see that picture of my sister in it, so I deactivated my account for about six months, I think."*

The participant has a responsibility to protect his hurma, in this case, his sister's reputation. A non-mahram was able to view a photo of his sister, which was in effect an exposure of his sister's awrah (in this case, her face). This is a shameful act, and one which brought embarrassment to the participant and his family.

In another interview a Saudi female (F2), who studied at a foreign university, explains that she needed to "*minimize her participation on Facebook*" to maintain an acceptable image back home. This came as a result of a scenario that not only affected her, but her family as well. She was chosen to appear in a promotional video for her university—she explains:

"*I was filmed* [unveiled]*…in a video…and I was tagged and some people from my hometown Googled me and saw it and didn't like it, and told my mom and she was pissed…All I remember is my sister calling and asking me to untag my name…But that just made me go back and check each picture on Facebook and check* [if it was] *private to make sure something like that will not happen again.*"

The function of tagging led to others from the student's community to become aware of the video. While she was proud of her participation, her family and those in her hometown saw her actions as unacceptable; she failed to adhere to khososyah. However, it is important to note that if a male were in the same situation, it would be a source of pride for his family, because the male face is not considered awrah; showing his face is acceptable, if not brag-worthy.

Similarly, (F3), Saudi female, explains that her participation on social media is affected by societal constraints:

*"I don't like that people could see pictures I am tagged in. Maybe it is fear of judgment or I didn't mature enough to justify it. Like if someone confronted me [about my photo] I wouldn't know what to say."*

F3 is expressing a fear of social stigma, which is a common theme in our data. Stigma and the maintenance of reputation have more serious consequences for women and their families in the GCC than they do on men. Therefore, users of social media are very aware of the consequences associated with any privacy violation. E.g., women often do not use photos of themselves in profile pictures because they are afraid that if a photo falls in the wrong hands someone could blackmail them with it. In the GCC, the acquisition and distribution of female photos is prohibited, as it is considered an attack on the family hurma [2,6,44].

Physical proximity is another factor that plays into what users post to their social media. (M4), a Saudi male participant who studied abroad, discusses his information disclosure practice on Facebook once he returned to Saudi:

*"When I was there [in the US] it was different (long pause)...I would post anything I want if it was proper and if it is on me only I wouldn't be ashamed of it, I would keep it public even. But here [in Saudi] I would not."*

When he was abroad, M4 had a different sense of privacy, one that made it personal, and in which only he faced consequences of information disclosure. Upon returning home, and being in close proximity to his family, he felt the need to modify his behavior regarding social media, and to conform to expectations.

These examples exemplify how users navigate social networking sites in a way to make them culturally appropriate, to make them "their own." Actions and behaviors that may seem innocent and/or innocuous to some can be viewed as intrusive by a citizen of the GCC. At the same time, the ways in which privacy is understood and built into systems does not meet their needs. The interpretations of privacy that tend to prevail in technology design can have significant consequences on users who do not subscribe to this same view. To illustrate, the first author asked M4 to imagine the following scenario:

You wake up and your account is public—everything you worked on in terms of privacy settings is gone.

*"I don't want to imagine that (nervous laughter)...I am actually going in and removing all these tags...sh\*t, may Allah forgive me, yeah that would hurt...hurt reputation, my relationship, relationships with people"*

Upon hearing this completely hypothetical situation, M4 became so nervous and fearful of the remote possibility that his privacy settings would be reset that he decided to remove all tags of himself in any photo. While this is an extreme example of a participant making a hasty decision about his Facebook account, it highlights an outlook that pervades throughout the GCC: the importance of *honor*.

**Honor**
Honor—*'ird*—is the single most important value that drives cultural norms and expectations in the GCC [16]. The ways in which honor manifest are subtle and nuanced, though it touches upon all aspects of daily life. It is the responsibility of each individual to uphold 'ird. By doing so, they protect their family and maintain good societal standing. As explained by [16]: "much of the organization of the Arab family can be understood in terms of *'ird* as a controlling value, legitimating the family structure and the 'modesty code' required of both men and women."

'ird is connected to the chastity of women, and has a sacred characteristic in GCC society [16]. The connection of 'ird to chastity is additionally highlighted by [28]: "[C]hastity and sexual modesty [are] also very highly valued. Applied primarily to women, these values [are] not only tied to family honor but [are] held to be a religious obligation as well." In this sense, chastity refers to more than a women's sexuality; it implies the need for them to display chaste behavior in all situations. In face-to-face settings, chaste behavior is implied by wearing appropriate garments, not interacting with non-mahrams, and not calling attention to oneself. In digital environments, women invoke chastity by using privacy settings, in addition to other tactics, to protect hurma, awrah, and above all, family 'ird.

'ird is the foundational attribute that motivates the beliefs and actions of our participants as they navigate various social media. Through our analysis, discussions of privacy can all be traced to participants' concerns regarding the maintenance of 'ird. So, while privacy is the word that is often used, the overarching theme or incentive bound up in protecting privacy in digital environments is directly related to the protection of 'ird.

The significance of 'ird is evident in this excerpt by (F8), a female Saudi. She explains: *"...here in Saudi privacy is forced on you. You have to be private....it is not a choice to be secluded...it is society telling you as a women to be private...because of religion, culture, society...and it is just expected of you to be private...if you post something that is a bit too public people will call you and be like 'why are you posting this? You are sharing too much information.'"*

F8 is referring to her responsibility to protect her family's reputation through the control of information sharing. In speaking about posting something "a bit too public," she is referring to her position as someone under surveillance by her family [44]. In effect, F8 has very little choice in what she can post; she can test limits, but this will likely result in admonishment, and her family will remind her that she must maintain privacy (i.e. protect their 'ird), at all times.

The earlier excerpt from M1—the man who deactivated his Facebook account as a result of photos of his sister being posted—is another example that highlights 'ird. If M1's sister shared pictures of herself in a hijab, it would not be an issue. However, a photo of her in an uncovered state was circulated, and a friend of M1's "liked" it, which brought shame to M1, and possibly damaged his 'ird.

In the following example, (M5) discusses his return to Saudi Arabia after being in the US. Though his friends are the same, there is now a greater risk associated with being closer to family and friends who will judge him for sharing:

*"As Saudis we know how to keep our private life private and what is suppose to be public, we still keep private."* First author: *"Like what?"* M5: *"Your name, your work, your mother's name...the worst example we have here...a lot of guys are ashamed of their mothers' name."*

In his reference to "mothers' names," M5 is referring to the significance of honor and shame as they manifest in everyday situations. In the GCC, men experience shame if their male friends know their mother's name; this custom is borne of tribal practices that persist today. A man's mahram women (mother, wife) are usually referred to by nicknames or in the abstract (e.g., "the family," "the mother of my children"). In this regard, men "consider the names of their female relatives a private part of their lives that they do not want to share with others" [4], i.e. female relatives' names are awrah, and must be protected. If names are known to other men, then 'ird is damaged. As this relates to digital behavior; the onus is on men to ensure that female relatives' names are not released nor spread.

**Gender**

Gender is a theme that is referenced, either blatantly or subtly, in every interview. Participants discussed privacy in terms of protecting awrah and maintaining hurma, and gender plays a significant role in these discussions.

*The Maintenance of Hurma*

When "privacy" is used by our participants to describe values, they are invoking the concepts hurma, awrah, and khososyah. For example, in this excerpt from an interview with a female Qatari (F1) who is speaking about Instagram, we learn why she has a private account:

*"...for us as Muslims...you have to be respectful...I had some incidents on my Instagram, like I had to actually block some people and I unfollowed a lot of people... because some of them are very rude...or inappropriate. That's why I made it private... like, my photos are private, and I don't trust [others] and I block, block, blocked them."*

F1 discusses how she manages her Instagram account to make it an acceptable, respectable place for her to interact with others while maintaining her values. She carefully guards her account, and does not allow outsiders to intrude upon it. The participant references Islam as the reason for her behavior, i.e. she seeks barakah while eliminating society's judgment. In effect, she is protecting her hurma, which in this case, is her social media account.

Societal expectations regarding gender roles also trump personal views. In this example, (M2), a Saudi male, explains his views on his sisters covering their hair when they travel outside Saudi Arabia. He is not strict about whether his sisters cover; they often remove their hijabs when they are with him. However, he finds it troubling if they were to publically post a photo of themselves on Facebook without a hijab. When asked if he would permit his sisters to post a photo of themselves uncovered, he says:

*"In general I will say no because of the culture; even though I don't mind it, but everyone else will mind it...it is culture backed by religion, it is both not one without the other. For example, when my sisters open their snaps [videos from Snapchat] and I am near them, they hide their phones so I don't see their friends without the veil."*

This quote illustrates the gravitas placed on conforming to expectations regarding what is permissible for females, as well as the obligation to protect others' hurma. M2's sisters are careful to protect their friends' awrah, in this case, images of themselves uncovered.

A gendered focus on protecting hurma is also discussed by F5 when she explains her process in moving from an all-female Instagram account to a mixed gendered account, and the actions associated with such a merger:

*"When I first started using Instagram I didn't have guys because I wanted to put pictures of me and my friends because some of them are wearing the hijab. Then I started having guys on my Instagram, so I removed [them]."*

In discussing photos, (F9), a Qatari female said that most will add pictures of flowers, or objects, but not their faces. If a female's photos are misused, it will bring disrespect to her family. The participant describes her fear:

*"...somebody will steal my pictures, and he will do some modification, and create a problem...and the boys, in the family will be very much embarrassed, and they will create a big issue for the girl."*

F9 is alluding to the enactment of khososyah by females in order to guard social media accounts, which they consider hurma, as they are places in which awrah may be exposed. Overall, we can think of the gender roles that manifest as an aspect of teamwork; men and women understand expectations—and test limits—as they use social media within the boundaries of their culture.

**CULTURALLY SENSITIVE DESIGN FOR PRIVACY**

Our discussions highlight the ways in which particular values are enacted through the use of digital technologies. Notably, these values are not necessarily understood— much less taken into consideration—when designers and developers create and/or augment technologies. Working within this vein of inquiry, we assert that a way to reduce cultural bias is to consider cross-cultural studies of privacy. By looking to research that examines various cultures and the value tensions associated with technology use, we give designers and developers the necessary tools to produce technologies that are appropriate for a broad audience.

That said, challenges persist in finding a systematic way to translate theoretical insights into technical design principles. Addressing such an endeavor is the goal of *value sensitive design (VSD)*, an approach that incorporates values into the process of designing artifacts [22]. VSD is a "theoretically grounded approach to the design of technology that accounts for human values in a principled and comprehensive manner throughout the design process." VSD also takes into account *indirect stakeholders,* i.e. those who do not interact with technology but who are affected by it [22]. Indirect stakeholders were prominent in our discussions, as in how they influence users' decisions about what information to share, and through which medium. Similar to VSD is the approach of Privacy by Design (PbD). PbD focuses solely on privacy, with the goal of "protecting privacy by embedding it into the design specifications of technologies, business practices, and physical infrastructures" [14]. Our data show how participants appropriate privacy settings to accommodate their needs. These findings, together with our knowledge of VSD and PbD, provide a foundation for principles regarding culturally sensitive technology development.

**Burgeoning Design Opportunities**

Designing around a notion of privacy bound up in honor, modesty and reputation is unchartered territory. Before we attempt to translate these complex and nuanced perceptions of privacy into design principles, we sought participants' opinions. In follow-up questions, we asked: "what would you do to improve <X>?" and "how do you imagine <X> changing to meet your needs?" Participants' responses resulted in several suggestions:

*"Private" as default setting.* Participants expressed lack of trust in some platforms because of incidents in which privacy were settings were reverted to 'default,' which was *public*. One way to avoid this is to make the default setting 'private.' This is in line with the PbD principle: "If an individual does nothing, their privacy still remains intact. No action is required on the part of the individual to protect their privacy—it is built into the system, by *default*." [14].

*"Real name" and "one account" policies.* Some participants confessed to managing more than one account on Facebook in particular—one for family and "judgmental" friends, and one for "open minded" friends. Some also said they use nicknames as profile names. We learned that although many refer to this practice as the use of "fake names," the names are in fact representations of themselves that they like to present to particular audiences; the digital platform is the space where they experience self-expression that cannot be had elsewhere. Policies that require users to maintain only one profile in their "real" names stymy this freedom that is highly valued by those who seek more autonomy than what their cultures permit.

We join others in calling for such policies to be more lenient and flexible [13,25]. In addition, we recommend the development of an interface that allows for management of multiple accounts. Due to the fear of context collapse, and the all too real possibility of harming one's family through information sharing, the ability to easily maintain more than one account would be a welcome feature by many.

*Gender sensitive recommendation system.* The example of Layla we highlight in the introduction demonstrates a situation where non-mahram were friend suggestions. However, the practice of gender segregation often extends to digital environments. So, we follow [7] in recommending a modification to *friend suggestion* algorithms to support same gender-and/or "kin relationship" suggestions. As a result, we anticipate a reduction in uncomfortable incidents for women in particular.

**LIMITATIONS AND FUTURE WORK**

While our research has uncovered many and varied findings, we do not claim that they hold for the entire GCC, nor for either country in particular. We interviewed thirty-three participants, all of whom are educated, English-speaking, and most of whom spent time abroad. We acknowledge that this group is likely more open-minded than many of their compatriots. However, this initial study provides the foundation for future work that will involve a larger, more diverse set of respondents, and ask more probing questions about privacy, social media use, and how to best design with values in mind. Overall, our preliminary findings support the need for further theorization that provides an enriched understanding of "privacy."

**DISCUSSION AND CONCLUDING REMARKS**

With the goal of supporting the HCI community in our mission towards more inclusive and culturally-sensitive design [22], we introduce the Islamic perspective of "privacy," and explain why it is critical to understand. *Hurma, awrah,* and the protection of them—*khososyah*—all come under the umbrella of maintaining honor. Our data point to how social media users in Saudi Arabia and Qatar interpret these concepts. While conventional interpretations of these actions indicate that our participants are discussing the enactment of "privacy," what is taking place involves more than what is typically viewed as "privacy management." The ways in which "privacy" is enacted among our participants goes beyond concerns for safety, security, and the ability to separate oneself from a larger group in a controlled manner. We are in fact observing adherence to Islamic teachings, maintenance of reputation, and the careful navigation of social media activity so as to preserve respect and modesty.

Contextually-grounded understandings of the ideologies and norms that guide social media use by a diverse population helps designers and researchers to not only create more inclusive, culturally-sensitive tools, but to understand how their own biases and values—while not knowingly harmful or problematic—create limitations that require considerable effort on behalf of some users to be able to safely and effectively use them.

This is the first contextually grounded study that explores how privacy is understood and enacted via social media by Gulf Arab Muslims. Our aim is to recognize the consequences of designing technologies with a limited view of "privacy." This research looks at what "privacy" is from a wider spectrum of users and investigates various foundations of the concept. We anticipate this will lead to design principles that will better guide developers and researchers going forward.